\documentclass[aps,prl,preprint,superscriptaddress]{revtex4}

\usepackage{graphicx,amsfonts,amsmath,amssymb,epsfig}
\input epsf
\graphicspath{{./figures}}
\begin{document}

\title{Deterministic loading of individual atoms to a high-finesse optical cavity}

\author{Kevin M. Fortier, Soo Y. Kim, Michael J. Gibbons, Peyman Ahmadi, and Michael S. Chapman}

\affiliation{School of Physics, Georgia Institute of Technology,
  Atlanta, GA 30332-0430 }

\begin{abstract}
Individual laser cooled atoms are delivered on demand from a single
atom magneto-optic trap to a high-finesse optical cavity using an
atom conveyor.  Strong coupling of the atom with the cavity field
allows simultaneous cooling and detection of individual atoms for
time scales exceeding $15$ s.  The single atom scatter rate is
studied as a function of probe-cavity detuning and probe Rabi
frequency, and the experimental results are in good agreement with
theoretical predictions.  We demonstrate the ability to manipulate
the position of a single atom relative to the cavity mode with
excellent control and reproducibility.
\end{abstract}
\pacs{32.80.Pj}

\maketitle

Cavity QED systems consisting of individual atoms localized in
high-finesse optical micro-cavities are a fundamental system in
quantum optics and have important applications to quantum
information processing \cite{Berman1994, Mabuchi2002}.  The
intrinsic entanglement of atoms within the cavity mode and the
cavity field provides a means to reversibly transfer quantum
information between matter and light, and the eventual leakage of a
photon from the well-defined mode of the cavity provides a means for
high-fidelity, long-distance quantum communication.  Although there
has been recent progress in generating probabilistic atom-photon
entanglements using free-space coupling of individual atoms
\cite{Blinov2004, Volz2006} and collective excitations of atomic
ensembles \cite{Matsukevich2004, Kuzmich2005, KimbleNature2005},
only cavity QED systems in the strong coupling regime can provide
both high entanglement probability and individual atomic qubits that
can be independently manipulated.

The unique capabilities of optical cavity QED systems require
controllably localizing individual atoms inside high-finesse, sub-mm
length optical cavities.  In the last decade, there has been
considerable progress in integrating laser cooled and trapped atoms
with optical cavity QED systems in the strong coupling regime
\cite{Boozer2006, Maunz2004, McKeever2004, McKeever2003,
NumannPRL2005, NumannNature2005, Ye1999, McKeeverScience2004,
Legero2004, Kuhn2002, Sauer2004}.  Individual atoms have been cooled
and stored in optical cavities for time spans exceeding a second
\cite{NumannNature2005, Ye1999}, and these advances have allowed
demonstration of single photon sources and studies of the cavity QED
system \cite{McKeeverScience2004, Legero2004, Kuhn2002}.

Previous experimental efforts have relied on probabilistic loading
of laser-cooled atoms into the cavity from free-falling atoms or
from an unknown number of atoms transferred from optical dipole
traps \cite{NumannNature2005, Sauer2004}. Eventually, practical
applications will require deterministic loading methods of single
atoms into the cavity.  In this Letter, we realize this goal by
incorporating a deterministically loaded atom conveyor
\cite{Kuher2001} that is used to deliver a precise number of atoms
into a high finesse resonator. We achieve storage times exceeding
$15$ s for atoms in the cavity with continuous cooling and
observation using cavity assisted cooling \cite{NumannNature2005,
Murr2006}. The atom-cavity interaction is studied as a function of
probe-cavity detuning and probe Rabi frequency, and the experimental
results are in good agreement with theoretical predictions.  We
demonstrate the ability to manipulate the position of a single atom
relative to the cavity mode with excellent control and
reproducibility.  The use of an atom conveyor was suggested in
\cite{Barrett1999} as a means to scale cavity QED interactions to
many atomic qubits, and the results of this work represent an
important step towards this goal.
\begin{figure}[h]
\begin{center}
\includegraphics[width=3in]{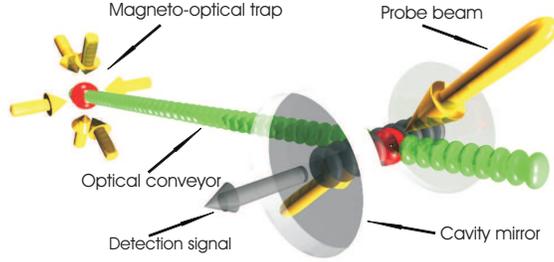}
\caption{A single atom MOT is formed $8.5$ mm away from the optical
cavity.  The atom is loaded into an atom conveyor and then
transported to the cavity mode.  Inside the cavity, the atom is
cooled via cavity-assisted cooling driven by counter-propagating
probe beams.} \label{fig1}
\end{center}
\end{figure}

A schematic of the experiment is illustrated in Fig.\thinspace
\ref{fig1}. A magneto-optical trap (MOT) of $^{87}$Rb atoms is
formed $8.5$ mm away from a high finesse optical cavity. The atoms
are counted in the MOT with single atom resolution.  Then they are
transported from the MOT to the cavity using an atom conveyor
\cite{Kuher2001} consisting of a 1-D optical lattice formed by two
independent counter-propagating laser beams from a fiber laser
operating at $ \lambda = 1064$ nm. The frequency of each lattice
beam is controlled by an acousto-optic modulator (AOM) and by
introducing a frequency difference between the counter-propagating
beams, the trapped atoms can be transported to the cavity.  This
lattice is focused at the cavity with a waist, $w_0 = 34~\mu$m and
provides a trapping potential of $U/k_B = 1$ mK with $4$ W optical
power per beam. The potential depth is only $100 ~\mu$K at the MOT
location ($8.5$ mm away) due to beam divergence. To increase the
loading probability of a single atom from the MOT into this shallow
trap, a separate loading lattice is employed. The loading lattice is
orthogonal to the conveyor axis and formed from a retro-reflected
beam focused to a $17 ~\mu$m waist at the MOT.  For a loading
lattice depth of $1$ mK, realized with 1 W of laser power, we
achieve $90\%$ efficiency in transferring atoms from the MOT to the
optical lattice.

The cavity is constructed from two super-polished concave mirrors
($r = 2.5$ cm) separated by $222 ~\mu$m with total measured losses
of $130$ ppm.  The relevant cavity QED parameters for this system
are $(g_0, \kappa, \gamma/2) = 2\pi (17, 7, 3)$ MHz where $g_0$,
$\kappa$, and $\gamma$, are the maximum atom-cavity coupling rate,
the cavity linewidth and the natural linewidth of the D2 line
$(5S_{1/2} \rightarrow 5P_{3/2})$ in $^{87}$Rb, respectively.  The
single atom cooperativity is $13.7$, putting the system in the
strong coupling regime.  The cavity length is actively stabilized to
the $F = 2 \rightarrow F' = 3$ transition of the $^{87}$Rb D2 line
($\lambda = 780$ nm) using an additional laser ($\lambda = 784$ nm)
that is tuned to resonance with a different longitudinal mode of the
cavity.
\begin{figure}[h]
\begin{center}
\includegraphics[width=3in]{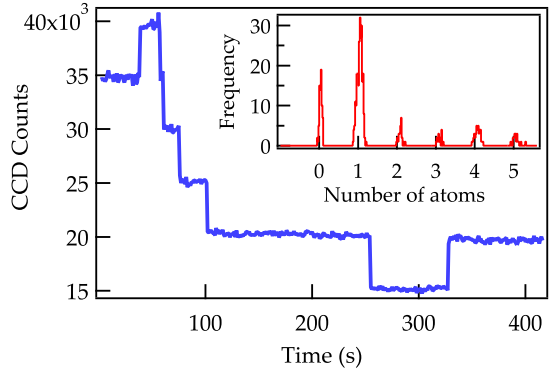}
\caption{The fluorescence signal collected from the high-gradient
MOT with an exposure time of 500 ms per data point. Discrete steps
indicate individual atoms captured or lost from the MOT. The inset
shows a histogram of the integrated fluorescence signal of 0 - 5
trapped atoms.} \label{fig2}
\end{center}
\end{figure}

The experiment begins by collecting and counting individual atoms in
the MOT.  To load single or small numbers of atoms, the MOT is
operated with magnetic field gradients of $~250$ G/cm to decrease
the loading volume.  This also provides tight confinement of the
atoms, localizing the trapped atoms to an area of approximately $25
\times 25 ~\mu \rm{m}^2$. The atoms are detected and counted by
measuring the fluorescence of the atoms from the MOT cooling beams.
This light is collected by a microscope objective (NA = $0.4$) and
focused onto an EMCCD (Andor IXon) camera. In Fig.\thinspace
\ref{fig2}, a typical time sequence of the MOT fluorescence is
shown. The discrete jumps of the observed fluorescence signal
correspond to individual atoms loading into or leaving the MOT. The
losses due to background gas collisions are sufficiently rare to
allow individual atoms to be trapped for more than $100$ s. By
exploiting the high quantum efficiency and the low noise of the
camera and by carefully minimizing background light (principally
stray scatter from the MOT beams), a single atom signal-to-noise
ratio of $>\!\!10\!:\!1$ is achieved with a $500$ ms integration
time.

Once the atoms are loaded into the MOT and counted, they are loaded
into the loading lattice.  This is accomplished by turning on both
the loading lattice and the conveyor lattice and, after a delay of
$100$ ms, turning off the MOT beams and magnetic field gradient. The
atom(s) are then transferred to the conveyor lattice by ramping the
loading lattice off in $75$ ms.  The atoms are transported to the
cavity at a velocity of $2.6$ cm/s by inducing a frequency
difference of $50$ kHz between the two counter-propagating beams of
the conveyor lattice. The atoms are brought to rest inside the
cavity mode by ramping the frequency difference to zero.  The
positioning reproducibility is estimated to be\thinspace $<\!\!10~
\mu$m, which corresponds to a relative position precision of
$10^{-3}$. The overall efficiency of transferring a known number of
atoms counted in the MOT to the cavity is $80\%$ for optimal
conditions.

Once the atom is inside the mode of the high finesse optical cavity,
it is continuously detected and cooled using cavity-assisted cooling
\cite{NumannNature2005, Vuletic2000, Vuletic2001}.  The atoms are
excited by two counter-propagating probe beams, and radiation
scattered from the atoms is re-emitted into the cavity mode and
subsequently detected by a photon counter as it leaks out the
cavity.  For positive cavity detunings with respect to the probe
beams ($\it{i.e,}~ \triangle_{\rm C} = \omega_{\rm c}- \omega_{\rm
p}
> 0$, where $\omega_{\rm c,\rm p}$ are the frequency of the cavity and the probe,
respectively) the photon absorbed by the atom from the probe has
lower energy than that emitted into the cavity mode, resulting in
net cooling of the atom.

The probe beams are oriented $45^\circ$ from the conveyor axis and
have a lin $\bot$ lin polarization configuration.  They are tuned
$21.5$ MHz below the $F = 2 \rightarrow F' = 3$ transition with a
Rabi frequency of $\Omega = (2\pi)~ 12$ MHz per beam and hence also
provide conventional Doppler cooling along the probe beam direction.
A hyperfine repumping laser beam co-propagates with these beams to
drive the $F = 1 \rightarrow F' = 2$ transition.  The emitted
photons from the cavity are detected with a single photon avalanche
photodiode (APD).

\begin{figure}[h]
\begin{center}
\includegraphics[width=3in]{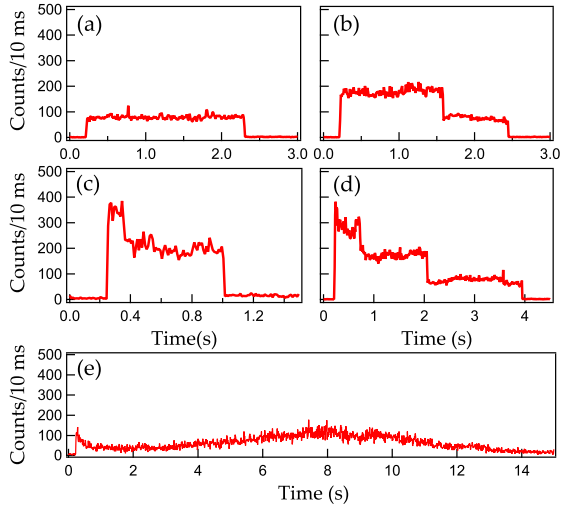}
\caption{Detected cavity emission signal vs. time.  In (a)-(d), the
cavity emission signal corresponds to 1-4 atoms, respectively,
initially loaded into the MOT and sub-sequently stored and detected
in the cavity.  In (e), storage of a single atom in the cavity for
15 s is shown.} \label{fig3}
\end{center}
\end{figure}

Typical cavity emission signals corresponding to deterministically
loaded atoms are shown in Fig.\thinspace \ref{fig3}(a-d) for
different numbers of atoms initially loaded in the MOT ($N_{\rm
atoms} = 1-4$, respectively). In each case, the probe is turned on
$250$ ms after the atom(s) are brought to rest inside the cavity.
The cavity emission signal is proportional to the number of atoms,
and corresponds to a detected count rate of $10$ counts/ms/atom. The
particular data shown in Fig.\thinspace \ref{fig3}(a-d) show atom
storage up to $4$ s, however the lifetime of the continuously cooled
atoms in the cavity varies significantly depending on the exact
experimental conditions and the number of atoms in the cavity. In
general, for $N_{\rm atoms} > 3$, the storage time is $<1$ s for the
experimental regime explored to date, while for $N_{\rm atoms} =
1-3$, storage times of $>\!\!15$ s have been observed.

For optimal experimental conditions, we have observed single atom
storage times exceeding $15$ s with good reproducibility.  A typical
example of such a signal is shown in Fig.\thinspace \ref{fig3}(e).
The drift in the single atom count rate in this trace is due to a
drifting frequency offset between the rf synthesizers that drive the
conveyor AOMs.  This results in a drift speed, $v \sim 0.5 ~\mu$m/s,
which eventually moves the atom out of the cavity mode.

According to the theoretical model of cavity-assisted cooling
developed in \cite{NumannNature2005, Murr2006}, the rate at which a
single atom scatters a photon into the cavity mode is given by:
\begin{equation}
R = 2 \kappa \frac{g^2}{\triangle^2_c + \kappa^2}
\frac{\Omega^2}{\triangle^2_a + \gamma^2},\label{Eq1}
\end{equation}
where $\triangle_a = \omega_0  - \omega_p + \triangle_S$ is the
detuning of probe beam with respect to the atom resonance,
$\omega_0$, including the Stark shift, $\triangle_S \sim (2 \pi) 83$
MHz due to the conveyor optical lattice. For the experimental
parameters of our system, Eq.\thinspace (\ref{Eq1}) predicts an
emission rate of $R = 2400$ photons/ms.  Our detection efficiency is
estimated to be $12.5\%$, including $50\%$ quantum efficiency of the
APD, $50\%$ in propagation losses from the cavity to the APD and
$50\%$ loss due to detection only one of the possible polarizations
of the light emitted by the cavity.  Accounting for this detection
efficiency, the predicted signal is $300$ counts/ms, which is a
factor of $\sim 30$ larger than measured in Fig.\thinspace
\ref{fig3}. This discrepancy varies from day-to-day; for optimal
conditions, single atom signals as high as $30$ counts/ms (10 times
smaller than predicted by Eq.\thinspace (\ref{Eq1})) have been
observed. Possible sources of signal discrepancy are
non-transmission losses in the cavity mirrors and a reduced
effective coupling due to the Zeeman structure of the atoms, and/or
varying Stark shifts that depends on the alignment of the conveyor
lattice.
\begin{figure}[h]
\begin{center}
\includegraphics[width=3in]{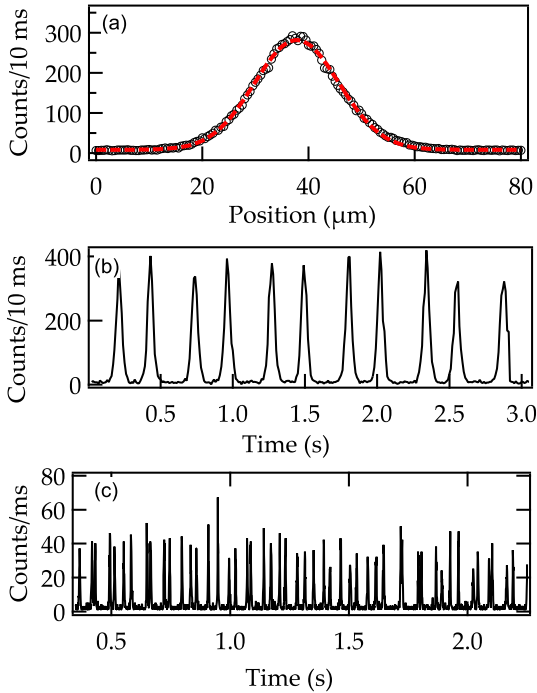}
\caption{In (a) an atom is swept across the cavity mode at a slow
speed, $v = 55~ \mu$m/s, to achieve the high resolution scan shown.
This data is the average of $17$ experimental runs which is fit to a
Gaussian function. (b) and (c) show a single atom swept across the
cavity mode $10$ and $75$ times with a speed of $440~ \mu$m/s and
$4.4$ mm/s, respectively.} \label{fig4}
\end{center}
\end{figure}

One of the advantages of the use of external fields to trap the atom
inside the cavity is that it allows control of the atom coupling via
the position dependence of the atom-cavity interaction strength.  In
Fig.\thinspace (\ref{fig4}), this control is exploited both to
investigate the position dependence of the coupling strength as well
as to repeatably move an atom in and out of the cavity mode.  For a
Fabry-Perot cavity, the coherent coupling rate of the
$\rm{TEM}_{00}$ Gaussian mode is given by $g({\bf r}) = g_0
\cos\thinspace (kz) \exp[-\rho^2/w^2]$ \cite{Berman1994}, written in
a cylindrical coordinate system with $z$ along the cavity axis and
where $w$ is the waist of the mode.  To study the dependence of the
coupling on the transverse coordinate, $\rho$, single atoms are
slowly moved through the cavity mode (with a speed of $55~ \mu$m/s)
while being continuously cooled and detected.  The single atom
signal vs. position are shown in Fig.\thinspace \ref{fig4}(a). The
data, which are an average of $17$ single atom runs, are fit well by
a Gaussian function as expected from Eq.\thinspace (\ref{Eq1}),
however, the measured waist ($w_0 = 16~ \mu$m), is $20\%$ smaller
than the waist calculated from the cavity geometry ($20~ \mu$m). The
source of this discrepancy is not known, but could be related to the
inadequacies of the 2-level model assumed in Eq.\thinspace
(\ref{Eq1}) or involve subtle interplays between the cooling rate
and the atom localization.

The atom conveyor allows for controllable and reversible
introduction of the atom into the cavity mode.  This is demonstrated
in Fig.\thinspace \ref{fig4}(b), which shows a single atom being
moved in and out of the cavity $10$ times. For this scan, the atomic
speed was $440~ \mu$m/s. Moving at faster velocities, we have
observed over $70$ passes of a single atom across the cavity mode.
Typical data for the faster scans is shown in Fig.\thinspace
\ref{fig4}(c).
\begin{figure}[h]
\begin{center}
\includegraphics[width=3in]{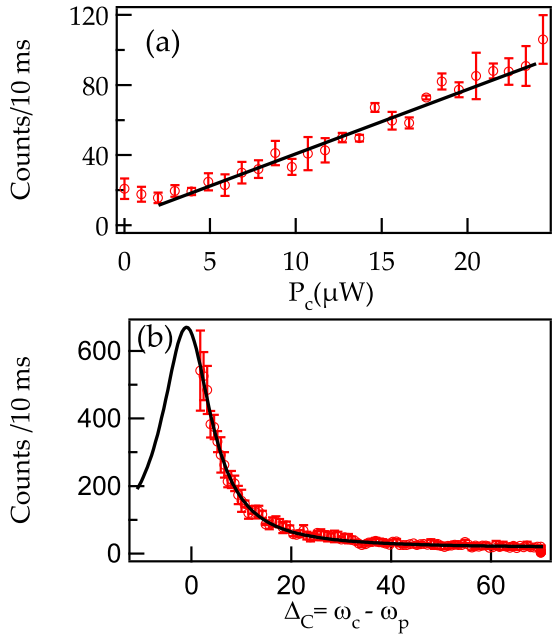}
\caption{(a)  Single atom scattering rate versus probe beam power.
The solid line is a linear fit to the data as expected from the
dependence of the scattering rate to the probe beam power. The first
three points are dark counts and are not included in the fit. (b)
The dependence of the single atom scattering rate with respect to
the cavity detuning $\triangle_C$.  The solid line is a fit to a
Lorentzian function.} \label{fig5}
\end{center}
\end{figure}

We have measured the single atom signal as a function of the power
of the probe beams and the detuning of the cavity as shown in
Fig.\thinspace \ref{fig5}.  For the data in Fig.\thinspace
\ref{fig5}(a), the power in the probe beams is linearly ramped from
$24$ nW to $24~ \mu$W in $250$ ms after a single atom has been
loaded into the cavity.  This corresponds to a variation in the Rabi
frequency of $\Omega = (2 \pi) 0.8 - 25$ MHz. For this data the
cavity detuning was $-12$ MHz and as expected from Eq.\thinspace
(\ref{Eq1}), the single atom signal is proportional to the power of
the cooling beams $(\propto \Omega^2)$. For the data in
Fig.\thinspace \ref{fig5}(b), a single atom is loaded in the cavity,
and the probe-cavity detuning, $\triangle_{\rm C} = \omega_{\rm c} -
\omega_{\rm p}$, is varied by detuning the cavity while holding the
frequency of the probe beam constant at $\omega_{\rm p} = \omega_0 -
21.5$ MHz.  Over the range of $\triangle_{\rm C}$ that is
investigated, the scattering rate shows a Lorentzian dependence on
$\triangle_{\rm C}$, with a line center and linewidth of zero MHz
and $6$ MHz (HWHM) in close agreement with the measured cavity
linewidth. This technique can only be used for investigating
positive values of $\triangle_{\rm C}$ because the negative values
of $\triangle_{\rm C}$ result in heating rather than cooling of the
atoms \cite{Murr2006} and lead to rapid loss of the trapped atoms.
In conclusion, we have demonstrated the deterministic delivery of
single atoms on demand to a high finesse optical cavity.  This was
accomplished by loading single atoms from a high gradient MOT into
an atom conveyor and subsequently transporting them into an optical
cavity. Employing cavity-assisted cooling, long storage times are
observed.  We also investigated the position dependence of the
single atom scattering rate and the dependence of this rate on the
power of the cooling beams and the detuning of the cavity.  The
results are in qualitative agreement with existing theoretical
calculations.  The successful integration of an atom conveyor with a
high finesse cavity opens the door to a scalable cavity QED based
quantum information processing system.  We would like to thank Paul
Griffin for stimulating discussions. This work was supported by the
NSF Grant PHY-0326315.


\begin{references}
\bibitem{Berman1994} P.R.\thinspace Berman, Cavity Quantum Electrodynamics (Academic Press,
Inc., San Diego, 1994).

\bibitem{Mabuchi2002} H.\thinspace Mabuchi, and A.C.\thinspace Doherty,
Science {\bf 298}, 1372 (2002).

\bibitem{Blinov2004} B.B.\thinspace Blinov {\it et al.}, Nature {\bf 428}, 153 (2004).

\bibitem{Volz2006} J.\thinspace Volz {\it et al.}, Phys.\thinspace Rev.\thinspace
Lett {\bf 96} 030404 (2006).

\bibitem{Matsukevich2004} D.N.\thinspace Matsukevich, and A.\thinspace Kuzmich,
Science {\bf 306}, 663 (2004).

\bibitem{Kuzmich2005} D.N.\thinspace Matsukevich {\it et al.},
Phys.\thinspace Rev.\thinspace Lett {\bf 95} 040405 (2005).

\bibitem{KimbleNature2005} C.W.\thinspace Chou {\it et al.},
 Nature {\bf 438}, 828 (2005).

\bibitem{Boozer2006} A.D.\thinspace Boozer {\it et al.},
Phys.\thinspace Rev.\thinspace Lett {\bf 97}, 83602 (2006).

\bibitem{NumannPRL2005} S.\thinspace Nu{\ss}mann {\it et al.},
Phys.\thinspace Rev.\thinspace Lett {\bf 95}, 173602 (2005).

\bibitem{NumannNature2005} S.\thinspace Nu{\ss}mann {\it et al.}, Nature Physics {\bf 1}, 122 (2005).

\bibitem{Maunz2004} P.\thinspace Maunz {\it et al.}, Nature {\bf 428}, 50 (2004).

\bibitem{McKeever2004} J.\thinspace McKeever {\it et al.},
Phys.\thinspace Rev.\thinspace Lett {\bf 93}, 143601 (2004).

\bibitem{McKeeverScience2004} J.\thinspace McKeever {\it et al.}, Science {\bf 303}, 1992 (2004).

\bibitem{Legero2004} T.\thinspace Legero {\it et al.}, Phys.\thinspace Rev.\thinspace Lett
{\bf 93}, 070503 (2004).

\bibitem{Sauer2004} J.A.\thinspace Sauer {\it et al.}, Phys.\thinspace Rev.\thinspace A.
{\bf 69}, 051804(R) (2004).

\bibitem{McKeever2003} J.\thinspace McKeever {\it et al.},
Phys.\thinspace Rev.\thinspace Lett {\bf 90}, 133602 (2003).

\bibitem{Kuhn2002} A.\thinspace Kuhn, M.\thinspace Hennrich, and G.\thinspace Rempe,
Phys.\thinspace Rev.\thinspace Lett {\bf 89}, 067901 (2002).

\bibitem{Ye1999} J.\thinspace Ye, D.W.\thinspace Vernooy, and H.J.\thinspace Kimble,
Phys.\thinspace Rev.\thinspace Lett {\bf 83}, 4987 (1999).

\bibitem{Kuher2001} S.\thinspace Kuhr {\it et al.}, Science {\bf 293}, 278 (2001).

\bibitem{Murr2006} K.\thinspace Murr {\it et al.}, Phys.\thinspace Rev.\thinspace A.
{\bf 73}, 063415 (2006).

\bibitem{Barrett1999} M.\thinspace Barrett, A.\thinspace Prasad, and M.S.\thinspace Chapman,
Bull. A. Phys. Soc. {\bf 13}, 6 (1999).

\bibitem{Vuletic2000} V.\thinspace Vuleti\'{c}, and S.\thinspace Chu, Phys.\thinspace
Rev.\thinspace Lett {\bf 84}, 3787 (2000).

\bibitem{Vuletic2001} Hilton\thinspace W.\thinspace Chan, Adam\thinspace T.\thinspace
Black, and Vladan\thinspace Vuleti\'{c}, Phys.\thinspace
Rev.\thinspace Lett {\bf 90}, 063003 (2003).
\end{references}
\end{document}